\documentclass[journal]{IEEEtran}

\usepackage{graphicx} \usepackage{bm} \usepackage{amsmath}
\usepackage{cite}
\usepackage{amssymb} 

\usepackage{amsmath,amsthm,amssymb,mathrsfs}
\usepackage{bm}
\usepackage{subfigure}

\ifCLASSINFOpdf
\else
\fi
\begin{document}

\title{Role of Magnetic Field in Self-Oscillation of Nanomagnet Excited by Spin Torque}
\author{Tomohiro~Taniguchi$^{\dagger}$,
        Shingo~Tamaru, 
        Hiroko~Arai,
        Sumito~Tsunegi, 
        Hitoshi~Kubota,
        and~Hiroshi~Imamura
        \\
        National Institute of Advanced Industrial Science and Technology (AIST), 
        Spintronics Research Center, \\
        Tsukuba, Ibaraki 305-8568, Japan
\thanks{${}^{\dagger}$Corresponding author. Email address: tomohiro-taniguchi@aist.go.jp}}

\maketitle

\begin{abstract}

The critical current of the self-oscillation 
of spin torque oscillator (STO) 
consisting of a perpendicularly magnetized free layer 
and an in-plane magnetized pinned layer was studied 
by solving the Landau-Lifshitz-Gilbert (LLG) equation. 
We found that the critical current diverged at certain field directions, 
indicating that the self-oscillation does not occur at these directions. 
It was also found that 
the sign of the critical current changed 
depending on the applied field direction. 

\end{abstract}


\IEEEpeerreviewmaketitle


\section{Introduction}
\label{sec:Introduction}

\IEEEPARstart{D}{irect} 
current applied to a magnetic tunnel junction (MTJ) 
exerts spin torque on the magnetization of the free layer. 
When the energy supplied by the spin torque balances with 
the energy dissipation due to the damping, 
the self-oscillation of the magnetization is realized. 
Spin torque oscillator (STO) [1]-[11] utilizing this self-oscillation 
is an important spintronics device 
applicable to 
microwave generators and 
recording heads of a high density hard disk drive (HDD) 
due to its small size, 
high emission power, 
and frequency tunability. 
Recent development of the experimental technique to enhance 
the perpendicular magnetic anisotropy of CoFeB 
by adding an MgO capping layer [12]-[14]
enabled us to fabricate 
STO consisting of a perpendicularly magnetized free layer 
and an in-plane magnetized pinned layer [15]-[18]. 
This type of STO in the presence of a large applied field (2-3 kOe) 
showed a large emission power ($\sim 0.5$ $\mu$W) 
with a narrow linewidth ($\sim 50$ MHz) [15]. 
On the other hand, 
the emission power 
in the absence of the applied field was 
of the order of $0.01$ $\mu$W [16]. 
These results indicate that the applied field plays a key role 
in the performance of STO. 
However, the previous works  only focused 
on the self-oscillation 
with the perpendicular field, 
while it is experimentally possible to 
apply the field in an arbitrary direction.


It is important for STO applications 
to clarify the dependence of the oscillation properties of STO 
on the applied field direction 
because of many reasons. 
For example, it was shown in Ref. [21] for an in-plane magnetized system 
that the oscillation frequency and the emission power 
could be controlled by changing the field direction. 
Also, when STO is used as a read-head sensor of HDD [22], 
the dipole field from a recording bit acts as an applied field 
whose direction depends on the bit information. 
The dependence of a critical current for the self-oscillation 
on the applied field direction is 
also interesting 
because the critical current determines the power consumption. 


In this paper, 
we study the dependence of the critical current 
of STO with a perpendicularly magnetized free layer 
and an in-plane magnetized pinned layer 
on the applied field direction. 
The critical current diverges 
when the field points to certain directions, 
indicating that the self-oscillation cannot be induced. 
This singularity arises from the energy balance 
between the work done by spin torque and 
the energy dissipation due to the damping. 
The field directions corresponding to the divergence 
depend on the perpendicular magnetic anisotropy, 
the applied field magnitude and direction, 
and the spin torque parameter. 
The result implies that 
the emission power of STO increases (decreases) significantly 
by tilting the magnetic field 
to the parallel (anti-parallel) direction 
of the pinned layer magnetization. 


\begin{figure}
  \centerline{\includegraphics[width=0.6\columnwidth]{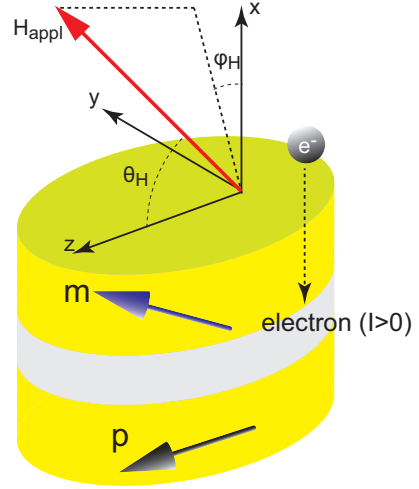}}
  \caption{
           Schematic view of the system. 
           The unit vectors pointing in the magnetization directions of 
           the free and the pinned layers are denoted as $\mathbf{m}$ and $\mathbf{p}$, respectively. 
           The $x$-axis is normal to the film-plane whereas the $z$-axis is parallel to $\mathbf{p}$. 
           The positive current corresponds to the electron flow from the free layer to the pinned layer. 
           The direction of the applied field $H_{\rm appl}$ is characterized by two angles $(\theta_{H},\varphi_{H})$, 
           where $\theta_{H}$ and $\varphi_{H}$ are angles from the $z$- and $x$-axes, respectively. 
  \vspace{-3.5ex}}
  \label{fig:fig1}
\end{figure}



This paper is organized as follows. 
In Sec. II, 
we linearize the Landau-Lifshitz-Gilbert (LLG) equation around 
the equilibrium state of the free layer. 
In Sec. III, 
the theoretical formula of the critical current is derived. 
The dependence of the critical current 
on the applied field direction is also studied. 
Section IV is devoted to the conclusion.



\section{LLG Equation}
\label{sec:LLG Equation}

The system we consider is schematically shown in Fig. \ref{fig:fig1}. 
We denote the unit vectors pointing in the directions of 
the magnetization of the free and the pinned layers 
as $\mathbf{m}$ and $\mathbf{p}$, respectively. 
The $z$-axis is parallel to $\mathbf{p}$ 
while the $x$-axis is normal to the film-plane. 
The current $I$ flows along the $x$-axis, 
where the positive current corresponds to the electron flow 
from the free layer to the pinned layer. 
The following calculations are based on the macrospin model, 
which works well as the volume of the free layer decreases. 
For our parameters shown below, 
where the radius and thickness of the free layer are 60 and 2 nm, respectively, 
the macrospin model well reproduced the experimental results 
such as the current dependence of the oscillation frequency [15]. 
On the other hand, when the radius and thickness of the free layer become 
larger than a few hundred and a few nanometers, respectively, 
the vortex state appears in the free layer [23]-[27]. 
Although the oscillation frequency of the vortex based STO is very low 
(typically, on the order of 0.1 GHz), 
a narrow linewidth on the order of sub MHz [27] 
is fascinating feature for practical applications. 
The magnetization dynamics of the macrospin is described by 
the LLG equation: 
\begin{equation}
  \frac{{\rm d}\mathbf{m}}{{\rm d}t}
  =
  -\gamma
  \mathbf{m}
  \times
  \mathbf{H}
  -
  \gamma
  H_{\rm s}
  \mathbf{m}
  \times
  \left(
    \mathbf{p}
    \times
    \mathbf{m}
  \right)
  +
  \alpha
  \mathbf{m}
  \times
  \frac{{\rm d}\mathbf{m}}{{\rm d} t}.
  \label{eq:LLG}
\end{equation}
The gyromagnetic ratio and Gilbert damping constant are denoted as 
$\gamma$ and $\alpha$, respectively. 
The magnetic field is defined by 
$\mathbf{H}=-\partial E/\partial (M \mathbf{m})$, 
where the energy density $E$ is 
\begin{equation}
  E
  =
  -MH_{\rm appl}
  \mathbf{m}
  \cdot
  \mathbf{n}_{H}
  -
  \frac{M(H_{\rm K}-4\pi M)}{2}
  \left(
    \mathbf{m}
    \cdot
    \mathbf{e}_{x}
  \right)^{2}. 
  \label{eq:energy}
\end{equation}
Here, $M$, $H_{\rm appl}$, 
$\mathbf{n}_{H}=(\sin\theta_{H}\cos\varphi_{H},\sin\theta_{H}\sin\varphi_{H},\cos\theta_{H})$, 
and $H_{\rm K}$ are 
the saturation magnetization, 
applied field magnitude, 
unit vector pointing in the applied field direction, 
and crystalline anisotropy field along the $x$-axis, respectively. 
The spin torque strength, $H_{\rm s}$ in Eq. (\ref{eq:LLG}), is [28]-[31]
\begin{equation}
  H_{\rm s}
  =
  \frac{\hbar \eta I}{2e (1 + \lambda \mathbf{m}\cdot\mathbf{p})MV}, 
  \label{eq:H_s}
\end{equation}
where $V$ is the volume of the free layer. 
Two dimensionless parameters, $\eta$ and $\lambda$, determine 
the magnitude of the spin polarization of the injected current 
and the dependence of the spin torque strength 
on the relative angle of the magnetizations, respectively. 


Because the parameter $\lambda$ plays a key role in the following discussions, 
explanations on its sign and value are mentioned in the following. 
The form of Eq. (\ref{eq:H_s}) is common 
for spin torque in not only MTJs but also giant magnetoresistive (GMR) systems [28]-[32], 
and the theoretical relation between $\lambda$ and the material parameters depend on the model. 
For example, Ref. [31] calculated 
the spin torque from the transfer matrix of an MTJ, 
and showed that $\lambda=\eta \eta^{\prime}$, 
where $\eta^{\prime}$ is the spin polarization of the free layer. 
The sign of $\lambda$ is positive (negative) 
when the MTJ shows the positive (negative) TMR. 
On the other hand, in the case of a GMR system, 
Ref. [30] calculated the spin torque 
from the ballistic spin current in a circuit, 
and showed that $\lambda=(\varLambda^{2}-1)/(\varLambda^{2}+1)$, 
where $\varLambda=\sqrt{R_{\rm F}/R_{\rm N}}$ depends on 
the resistances of the ferromagnetic (F) electrode and the nonmagnetic (N) spacer. 
Depending on the ratio $R_{\rm F}/R_{\rm N}$, 
$\lambda$ can be both positive and negative. 



\begin{figure}
  \centerline{\includegraphics[width=0.7\columnwidth]{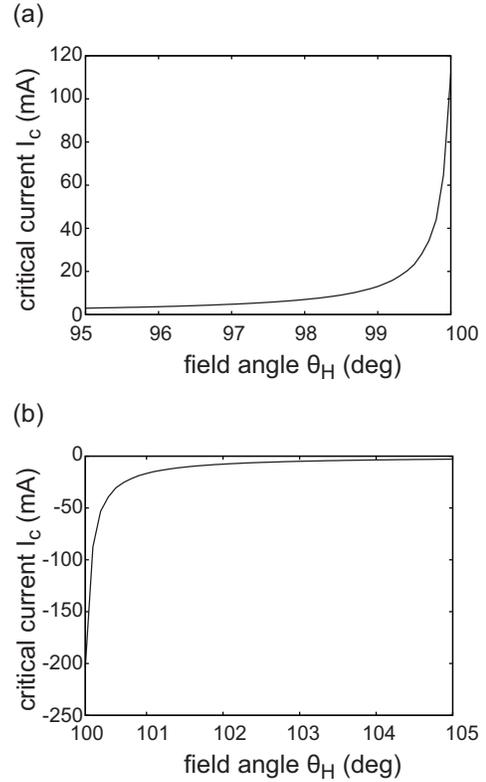}}
  \caption{
           Dependences of the critical current 
           on the applied field angle $\theta_{H}$ for 
           (a) $95^{\circ} \le \theta_{H} \le 100^{\circ}$ 
           and 
           (b) $100^{\circ} \le \theta_{H} \le 105^{\circ}$, 
           where $\varphi_{H}=0$. 
  \vspace{-3.5ex}}
  \label{fig:fig2}
\end{figure}



Before applying the current, 
the magnetization points to the equilibrium direction 
corresponding to the minimum of the energy density $E$. 
In the spherical coordinate, 
the polar and azimuth angles $(\theta,\varphi)$ 
of the equilibrium direction satisfy 
\begin{equation}
\begin{split}
  &
  H_{\rm appl}
  \left[
    \sin\theta_{H}
    \cos\theta
    \cos(\varphi_{H}-\varphi)
    -
    \cos\theta_{H}
    \sin\theta
  \right]
\\
  &+
  \left(
    H_{\rm K}
    -
    4\pi M
  \right)
  \sin\theta
  \cos\theta
  \cos^{2}\varphi
  =
  0,
  \label{eq:condition_eq_1}
\end{split}
\end{equation}
\begin{equation}
\begin{split}
  &
  H_{\rm appl}
  \sin\theta_{H}
  \sin\theta
  \sin(\varphi_{H}-\varphi)
\\
  &-
  \left(
    H_{\rm K}
    -
    4\pi M
  \right)
  \sin^{2}\theta
  \sin\varphi
  \cos\varphi
  =
  0.
  \label{eq:condition_eq_2}
\end{split}
\end{equation}
We introduce a new coordinate $XYZ$ 
in which the $Z$-axis points to $(\theta,\varphi)$ direction. 
The transformation matrix 
from the $xyz$-coordinate to the $XYZ$-coordinate is given by 
\begin{equation}
  \mathsf{R}
  =
  \begin{pmatrix}
    \cos\theta & 0 & -\sin\theta \\
    0 & 1 & 0 \\
    \sin\theta & 0 & \cos\theta
  \end{pmatrix}
  \begin{pmatrix}
    \cos\varphi & \sin\varphi & 0 \\
    -\sin\varphi & \cos\varphi & 0 \\
    0 & 0 & 1
  \end{pmatrix}.
\end{equation}
The magnetic field in the $XYZ$-coordinate is 
$\mathbf{H}=(H_{XX}m_{X}+H_{XY}m_{Y},H_{YX}m_{X}+H_{YY}m_{Y},H_{ZX}m_{X}+H_{ZY}m_{Y}+H_{ZZ})$, 
where $H_{ij}$ are the $i$-components of $\mathbf{H}$ 
proportional to $m_{j}$ [33],[34]. 
The explicit forms of $H_{ij}$ are 
\begin{equation}
  H_{XX}
  =
  \left(
    H_{\rm K}
    -
    4\pi M
  \right)
  \cos^{2}\theta
  \cos^{2}\varphi,
  \label{eq:H_XX}
\end{equation}
\begin{equation}
  H_{YY}
  =
  \left(
    H_{\rm K}
    -
    4\pi M
  \right)
  \sin^{2}\varphi,
  \label{eq:H_YY}
\end{equation}
\begin{equation}
  H_{XY}
  =
  H_{YX}
  =
  -\left(
    H_{\rm K}
    -
    4\pi M
  \right)
  \cos\theta
  \sin\varphi
  \cos\varphi,
  \label{eq:H_XY}
\end{equation}
\begin{equation}
  H_{ZX}
  =
  \left(
    H_{\rm K}
    -
    4\pi M
  \right)
  \sin\theta
  \cos\theta
  \cos^{2}\varphi,
  \label{eq:H_ZX}
\end{equation}
\begin{equation}
  H_{ZY}
  =
  -\left(
    H_{\rm K}
    -
    4\pi M
  \right)
  \sin\theta
  \sin\varphi
  \cos\varphi,
  \label{eq:H_ZY}
\end{equation}
\begin{equation}
\begin{split}
  H_{ZZ}
  =&
  H_{\rm appl}
  \left[
    \sin\theta_{H}
    \sin\theta
    \cos(\varphi-\varphi_{H})
    +
    \cos\theta_{H}
    \cos\theta
  \right]
\\
  &+
  \left(
    H_{\rm K}
    -
    4\pi M
  \right)
  \sin^{2}\theta
  \cos^{2}\varphi. 
  \label{eq:H_ZZ}
\end{split}
\end{equation}
Up to the first order of $m_{X}$ and $m_{Y}$, 
the LLG equation can be linearized as 
\begin{equation}
  \frac{1}{\gamma^{\prime}}
  \frac{{\rm d}}{{\rm d}t}
  \begin{pmatrix}
    m_{X} \\
    m_{Y} 
  \end{pmatrix}
  +
  \mathsf{M}
  \begin{pmatrix}
    m_{X} \\
    m_{Y} 
  \end{pmatrix}
  =
  \begin{pmatrix}
    H_{\rm s0}\sin\theta \\
    0 
  \end{pmatrix},
  \label{eq:linear_LLG}
\end{equation}
where 
the components of the $2 \times 2$ matrix, $\mathsf{M}$, are given by 
\begin{equation}
\begin{split}
  \mathsf{M}_{1,1}
  =&
  -H_{YX}
  +
  \alpha
  \left(
    H_{ZZ}
    -
    H_{XX}
  \right)
\\
  &-
  H_{\rm s0}
  \left(
    \cos\theta
    +
    \tilde{\lambda}
    \sin\theta
  \right),
  \label{eq:M_11}
\end{split}
\end{equation}
\begin{equation}
  \mathsf{M}_{1,2}
  =
  \left(
    H_{ZZ}
    -
    H_{YY}
  \right)
  -
  \alpha
  H_{XY},
  \label{eq:M_12}
\end{equation}
\begin{equation}
  \mathsf{M}_{2,1}
  =
  -\left(
    H_{ZZ}
    -
    H_{XX}
  \right)
  -
  \alpha
  H_{YX},
  \label{eq:M_21}
\end{equation}
\begin{equation}
  \mathsf{M}_{2,2}
  =
  H_{XY}
  +
  \alpha
  \left(
    H_{ZZ}
    -
    H_{YY}
  \right)
  -
  H_{\rm s0}
  \cos\theta,
  \label{eq:M_22}
\end{equation}
where $H_{\rm s0}=\hbar \eta I/[2e(1 + \lambda \cos\theta)MV]$ and 
$\tilde{\lambda}$ is given by 
\begin{equation}
  \tilde{\lambda}
  =
  \frac{\lambda \sin\theta}{1+\lambda \cos\theta}.
  \label{eq:lambda_tilde}
\end{equation}


\section{Critical Current}
\label{sec:Critical Current}

Equation (\ref{eq:linear_LLG}) indicates that 
the time-evolutions of $m_{X}$ and $m_{Y}$ are described by ${\rm exp}\{\gamma[\pm{\rm i}\sqrt{{\rm det}[\mathsf{M}]-({\rm Tr}[\mathsf{M}]/2)^{2}}-{\rm Tr}[\mathsf{M}]/2]t\}$, 
where ${\det}[\mathsf{M}]$ and ${\rm Tr}[\mathsf{M}]$ are the determinant and trace of the matrix $\mathsf{M}$, respectively. 
The imaginary part of the exponent determines the oscillation frequency of $m_{X}$ and $m_{Y}$, 
which is identical to the ferromagnetic resonance frequency in the limit of $\alpha \to 0$ and $H_{\rm s} \to 0$. 
On the other hand, when the real part of the exponent ($-\gamma{\rm Tr}[\mathsf{M}]t/2$) is positive (negative), 
the amplitude of $m_{X}$ and $m_{Y}$ increases (decreases) with the time increases. 
Therefore, the critical current is determined by the condition ${\rm Tr}[\mathsf{M}]=0$, 
and is given by 
\begin{equation}
  I_{\rm c}
  =
  \frac{2\alpha e(1+\lambda \cos\theta) MV}{\hbar \eta (2 \cos\theta + \tilde{\lambda} \sin\theta)}
  \left(
    2 H_{ZZ}
    -
    H_{XX}
    -
    H_{YY} 
  \right).
  \label{eq:Ic}
\end{equation}
When the applied field points to the perpendicular direction ($\mathbf{n}_{H}=\mathbf{e}_{x}$) 
and $H_{\rm K} > 4\pi M$, 
the equilibrium direction is $(\theta,\varphi)=(\pi/2,0)$. 
In this case, the critical current is [18] 
\begin{equation}
  \lim_{(\theta,\varphi)\to(90^{\circ},0)}
  I_{\rm c}
  =
  \frac{4 \alpha eMV}{\hbar \eta \lambda}
  \left(
    H_{\rm appl}
    +
    H_{\rm K}
    -
    4\pi M
  \right).
  \label{eq:Ic_perp}
\end{equation}



\begin{figure}
  \centerline{\includegraphics[width=1.0\columnwidth]{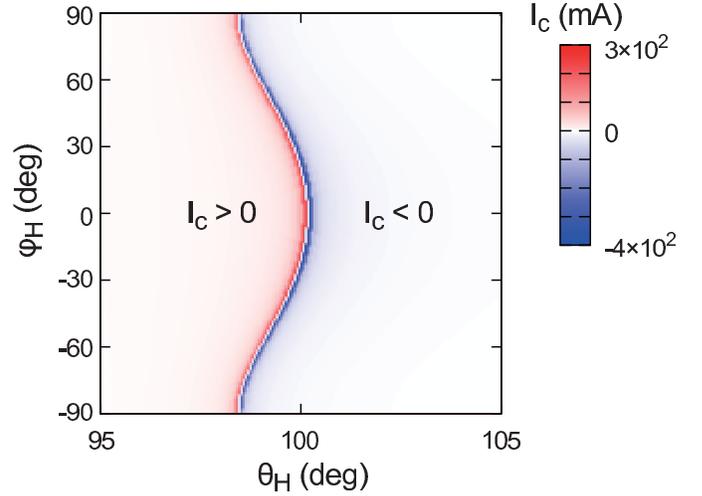}}
  \caption{
           Map of the critical current $I_{\rm c}$ 
           in $(\theta_{H},\varphi_{H})$ space. 
           Along the red and blue lines, 
           the critical current diverges. 
           The sign of the critical current is positive (negative) 
           in the left (right) side from 
           the line of the divergence. 
  \vspace{-3.5ex}}
  \label{fig:fig3}
\end{figure}


Figures \ref{fig:fig2} (a) and (b) show 
the dependences of the critical current $I_{\rm c}$ 
on the applied field angle $\theta_{H}$ 
for (a) $95^{\circ} \le \theta_{H} \le 100^{\circ}$ 
and (b) $100^{\circ} \le \theta_{H} \le 105^{\circ}$, respectively, 
in which $\varphi_{H}=0$. 
Also, in Fig. \ref{fig:fig3}, 
$I_{\rm c}$ in $(\theta_{H},\varphi_{H})$ space is shown, 
in which $95^{\circ} \le \theta_{H} \le 105^{\circ}$ 
and $-90 \le \varphi_{H} \le 90^{\circ}$. 
The values of the parameters are 
$M=1448$ emu/c.c., 
$H_{\rm K}=18.6$ kOe, 
$H_{\rm appl}=2$ kOe, 
$V=\pi \times 60 \times 60 \times 2$ nm${}^{3}$ 
$\eta=0.54$, 
$\lambda=\eta^{2}$, 
$\gamma=1.732 \times 10^{7}$ rad/(Oe$\cdot$s), 
and $\alpha=0.005$, respectively, 
which are estimated from the experiments [15],[35],[36]. 
Two important conclusions are obtained from Eq. (\ref{eq:Ic}), 
Figs. \ref{fig:fig2} (a), \ref{fig:fig2} (b), 
and \ref{fig:fig3}. 


The first conclusion is that the critical current diverges 
at certain field directions $(\tilde{\theta}_{H},\tilde{\varphi}_{H})$ 
as $\lim_{(\theta_{H},\varphi_{H}) \to (\tilde{\theta}_{H}-0,\tilde{\varphi}_{H})} I_{\rm c} = + \infty$ 
and $\lim_{(\theta_{H},\varphi_{H}) \to (\tilde{\theta}_{H}+0,\tilde{\varphi}_{H})} I_{\rm c} = - \infty$. 
For example, 
the critical current diverges near $\tilde{\theta}_{H} \sim 100^{\circ}$ 
for $\varphi_{H}=0$, 
as shown in Figs. \ref{fig:fig2} (a) and \ref{fig:fig2} (b). 
At $(\tilde{\theta}_{H},\tilde{\varphi}_{H})$, 
the condition 
\begin{equation}
  2 \cos\theta 
  +
  \frac{\lambda \sin^{2}\theta}{1 + \lambda \cos\theta}
  =
  0,
  \label{eq:condition_Ic}
\end{equation}
is satisfied, 
which means that the denominator of Eq. (\ref{eq:Ic}) is zero. 
It should be noted that 
Eq. (\ref{eq:condition_Ic}) depends on 
not only $\lambda$ 
but also the perpendicular magnetic anisotropy and 
applied field magnitude and direction 
through Eqs. (\ref{eq:condition_eq_1}) and (\ref{eq:condition_eq_2}). 


The divergence of the critical current means that 
the self-oscillation cannot be induced at the field direction 
satisfying Eq. (\ref{eq:condition_Ic}). 
It should be noted that 
the self-oscillation is realized 
when the energy supplied by the spin torque balances with 
the energy dissipation due to the damping. 
However, for example, when $\lambda=0$ and 
the equilibrium direction of the magnetization is 
perpendicular to the film plane 
($(\theta,\varphi)=(90^{\circ},0)$), 
the total energy supplied by the spin torque 
during a precession of the magnetization around the perpendicular axis is zero, 
as mentioned in Ref. [18]. 
Then, the critical current, Eq. (\ref{eq:Ic_perp}), in the limit of $\lambda \to 0$ diverges, 
and a steady oscillation does not occur by the spin torque. 
Equation (\ref{eq:condition_Ic}) should be regarded as 
the generalized condition for an arbitrary pointing field, 
which as a special case corresponds to $\lambda=0$ 
for the perpendicular field, 
i.e., the energy supplied by the spin torque is zero 
when the field, anisotropy, and $\lambda$ satisfy Eq. (\ref{eq:condition_Ic}). 
A significant reduction of the emission power is expected in experiments 
when the field points to the direction $(\tilde{\theta}_{H},\tilde{\varphi}_{H})$. 


The second conclusion is that 
the sign of the critical current changes 
from positive to negative 
when crossing the line of the divergence of $I_{\rm c}$ 
in Fig. \ref{fig:fig3} 
from left to right: 
see also Figs. \ref{fig:fig2} (a) and (b). 
To understand this point, 
it is convenient to consider the case 
in which a large field ($H_{\rm appl} \gg | H_{\rm K}-4\pi M |$) 
points to the parallel (P) or anti-parallel (AP) direction 
of $\mathbf{p}$. 
Then, the system can be regarded as the in-plane magnetized system. 
It is known that 
the sign of the critical current for the switching from P to AP state 
in the in-plane magnetized system is 
opposite to that for the switching from AP to P state [37]. 
In our definition, 
the sign of the critical current from P (AP) to AP (P) state 
is positive (negative), 
which is consistent with the sign of $I_{\rm c}$ 
shown in Figs. \ref{fig:fig2} (a) and (b). 
One of the important conclusions in Ref. [15] is 
that only the positive current can exert the self-oscillation of the magnetization 
in this type of STO. 
However, the present result suggests that 
the negative current can also exert the self-oscillation 
when the field tilts to the anti-parallel direction of $\mathbf{p}$. 


Finally, let us briefly discuss the relation between 
the present result and the STO applications. 
In STO applications, 
it is desirable to obtain a large emission power 
by using a low bias current. 
The emission power depends on 
the precession amplitude $m_{z}$. 
The previous experiments mainly focused on 
the effect of the perpendicular field ($\mathbf{n}_{H}=\mathbf{e}_{x}$) [15]. 
On the other hand, the present calculation shown in Fig. \ref{fig:fig2} (a) implies that 
the emission power increases significantly 
by slightly tilting the magnetic field 
from the perpendicular direction 
to the parallel direction of $\mathbf{p}$ 
because $m_{z}$ with a large amplitude at the low current is expected 
due to the decrease of the critical current $I_{\rm c}$. 
However, a further tilting of the magnetic field 
leads to an decrease of the oscillation amplitude 
because the oscillation is limited to the region $m_{z}>0$. 
Also, a complex magnetization dynamics may happen 
due to a large difference of the directions 
between the applied field and the anisotropy field. 
These considerations imply an existence of 
an optimum direction of the applied field for a high emission power, 
which will be an important work to pursue in future. 



\section{Conclusions}
\label{sec:Conclusions}

In conclusion, 
the critical current of STO 
consisting of a perpendicularly magnetized free layer 
and an in-plane magnetized pinned layer 
in the presence of an applied field 
pointing to an arbitrary direction was calculated. 
The critical current diverged at certain field directions, 
which meant that the self-oscillation could not be induced by the spin torque. 
The field direction corresponding to this singularity 
depended on the perpendicular magnetic anisotropy, 
the applied field magnitude and the direction, 
and the spin torque parameter. 
The divergence arose from the fact that 
the spin torque could not supply the energy to the free layer 
at these directions. 
It was also found that 
the sign of the critical current changed 
depending on the field direction. 


\section*{Acknowledgment}

The authors would like to acknowledge 
H. Maehara, A. Emura, T. Yorozu, M. Konoto, K. Yakushiji, T. Nozaki, A. Fukushima, K. Ando, and S. Yuasa. 
This work was supported by JSPS KAKENHI Number 23226001. 


\ifCLASSOPTIONcaptionsoff
  \newpage
\fi



\end{document}